\documentclass[prd,twocolumn,showpacs]{revtex4-1}

\usepackage{amsmath}
\usepackage{bm}
\usepackage{graphicx}
\usepackage{epstopdf}
\usepackage{hyperref}
\usepackage{soul}
\usepackage{color}

\enlargethispage{\baselineskip}

\hypersetup{
     colorlinks   = true,
     citecolor    = blue
}

\newcommand{\vf}{\varphi}
\newcommand{\dvf}{\dot{\varphi}}
\newcommand{\ddvf}{\ddot{\varphi}}
\newcommand{\df}{\dot{\phi}}
\newcommand{\ddf}{\ddot{\phi}}
\newcommand{\ca}{\hat{\bf a}_{\bf k}}
\newcommand{\Rk}{\mathcal{R}_k}
\newcommand{\Pk}{\Delta_{\mathcal{R}}(k_*)}
\newcommand{\PT}{\Delta_{\mathcal{T}}(k_*)}

\newcommand{\du}{\dot{u}}

\newcommand{\bz}{\bar{\zeta}_1}

\begin{document}

\newcommand{\FIRSTAFF}{\affiliation{The Oskar Klein Centre for Cosmoparticle Physics,
	Department of Physics,
	Stockholm University,
	AlbaNova,
	10691 Stockholm,
	Sweden}}
\newcommand{\SECONDAFF}{\affiliation{Nordita,
	KTH Royal Institute of Technology and Stockholm University,
	Roslagstullsbacken 23,
	10691 Stockholm,
	Sweden}}
\newcommand{\THIRDAFF}{\affiliation{University of Helsinki,
	P.O. Box 64, Helsinki, FI-00014
	Finland.}}

\title{Inflating without a flat potential: Viscous inflation}

\author{Luca Visinelli}
\email[Electronic address: ]{luca.visinelli@fysik.su.se}
\FIRSTAFF
\SECONDAFF
\THIRDAFF
\date{\today}

\begin{abstract}
Bulk viscosity leads to negative pressure, which is a key ingredient for successful inflation. We build an inflationary model where the slow-roll of the inflaton field is driven by a viscous component instead of the flat potential commonly used. Since viscosity does not contribute to the energy density of the Universe, the Hubble rate depends only on the kinetic energy of the inflaton. The power spectrum is almost scale-invariant, with a tilt depending on the slow-roll parameters defined within the model.
\end{abstract}
\maketitle

\section{Introduction} \label{sec1}

Inflation~\cite{kazanas, starobinsky, guth, sato, mukhanov, albrecht, linde_1982, linde_1983} provides a mechanism for explaining the observed flatness, homogeneity, and the lack of relic monopoles in the early Universe~\cite{mukhanov_rev, linde_book, kolb_book, bergstrom_book, weinberg_book}, as well as generating the inhomogeneities observed in the cosmic microwave background radiation~\cite{guth_pi, hawking, starobinsky1, bardeen1}. Realistic microphysical models of inflation, in which the expansion of the universe is governed by the energy density of an inflaton field $\vf$, require the inflaton potential $U = U(\vf)$ to be approximately flat in order for successful inflation to occur~\cite{adams_1991}. The inflationary stage lasts until the inflaton field begins to oscillate around the minimum of its potential, decaying into lighter degrees of freedom~\cite{dolgov_1982, abbott_1982, dolgov_1990, trashen_1990, kofman_1994}.

Bulk viscosity, which phenomenologically acts as a negative pressure term, might arise in the fluid description of a particle ensemble through various mechanisms like inter-particle interaction~\cite{landau_book, huang_book} or the decay of particles within the fluid~\cite{zeldovich1970, murphy1973, hu1982}. Dissipative processes, described by the relativistic theory of viscosity~\cite{hwang1990, hiscock, pavon_1991}, might have driven inflation, as discussed within the theory of the ``standard'' inflation theory~\cite{diosi_1986, waga_1986, barrow1986, barrow1987, barrow1988, zimdahl, pacher1987, zimdahl, zakari, maartens, zimdahl1996a, zimdahl1996b, zimdahl2000, maartens1, giovannini2005a, giovannini2005b, cataldo2006}, warm inflation~\cite{delcampo2007, delcampo2010, basterogil2011, basterogil2012, basterogil2014, visinelli2015}, or the present observed accelerated expansion of the Universe~\cite{bogdanos, avelino2009, gagnon2011, brevik, piattella}. Viable mechanisms for the generation of bulk viscosity from a field description have been suggested~\cite{zimdahl2000b, zimdahl2001, wilson2007, mathews2008, gagnon2011, basterogil2011, basterogil2012, basterogil2014}. 

In the literature, there has been some debate whether bulk viscosity can drive inflation. The authors in Refs.~\cite{barrow1986, barrow1987, barrow1988} add a viscous term to a relativistic fluid to obtain an inflationary stage, but see Ref.~\cite{pacher1987}. Inflation cannot be supported in a cosmology where the expansion is dominated by a fluid with a perfect gas equation of state~\cite{hiscock}, but the outcome changes if a different equation of state is used~\cite{pavon_1991, zimdahl, zakari, maartens, zimdahl1996a, zimdahl1996b, zimdahl2000, maartens1}, or in a two-fluid model~\cite{bamba2015, brevik2015}.

In this Letter we point out that a field whose energy density is mainly in the kinetic component can drive an inflationary period, if a bulk viscosity term $\zeta$ is present. Splitting the viscosity into a constant $\zeta_0$ and a time-varying component $\zeta_1$, we find that the constant part is related to the Hubble scale during inflation, while the time-depending part controls the slow rolling of the field. The scalar power spectrum nearly scale-invariant, with the tilt gauged by two slow-roll parameters. Thus, the inclusion of a non-equilibrium viscous term leads to the same phenomenological results as those obtained with an approximately flat inflaton potential $U$, which we do not include in the treatment. We revise perturbations and results for the standard inflation in Secs~\ref{sec2} and~\ref{sec3}, respectively, since most expression are later used in Sec.~\ref{sec4} where viscous inflation is presented.

\section{Cosmological perturbations during inflation} \label{sec2}

In an expanding universe, a fluid of energy density $\rho$ and pressure $p$ satisfies the energy conservation relation
\begin{equation} \label{energy_conservation}
\dot{\rho} + 3H(p + \rho) = 0,
\end{equation}
where the Hubble rate $H = \dot{a}/a$ is given in terms of the scale factor $a = a(t)$ and its time derivative $\dot{a}$ with respect to cosmic time $t$. Inflation is achieved whenever $\ddot{a} \geq 0$, or $H^2 + \dot{H} \geq 0$. The Hubble rate and its time derivative follow
\begin{equation} \label{hubble_rate}
H^2 = \frac{8\pi\,G}{3}\,\rho \quad \hbox{and} \quad \dot{H} = -4\pi\,G\,\left(p+\rho\right),
\end{equation}
where the expression for $\dot{H}$ follows from using Eq.~\eqref{energy_conservation}. Varying the density and pressure fields as
\begin{equation}
\rho \to \rho + \delta\rho \quad\hbox{and}\quad p \to p + \delta p,
\end{equation}
the expressions for the first-order perturbation terms are
\begin{eqnarray} \label{perturbation_energy_malik}
\delta \dot{\rho} \!+\! 3H\,\left(\delta \rho \!+\! \delta p\right) \!-\! 3\left(p \!+\! \rho\right)\,\dot{\Psi} \!+\! \frac{\nabla^2}{a^2}\,\left(p \!+\! \rho\right)\,\delta u &=& 0,\\
\dot{\Psi} + H\,\Psi + 4\pi\,G \,\left(p+\rho\right)\, \delta u &=& 0,
\end{eqnarray}
where $\Psi$ is the perturbation in the gravitational field and $\delta u$ is the total covariant velocity perturbation. Assuming a scale of inflation $H_I \ll M_{\rm Pl}$ allows us to neglect the perturbations in the gravitational field in Eq.~\eqref{perturbation_energy_malik}, or~\cite{weinberg_book}
\begin{equation} \label{perturbation_energy_malik1}
\delta \dot{\rho} + 3H\,\left(\delta \rho + \delta p\right) + \frac{\nabla^2}{a^2}\,\left(p+\rho\right)\,\delta u = 0.
\end{equation}

\section{Standard inflationary model} \label{sec3}

In the standard theory of inflation~\cite{kazanas, starobinsky, guth, sato, mukhanov, albrecht, linde_1982, linde_1983, mukhanov_rev, linde_book, kolb_book, bergstrom_book, weinberg_book, guth_pi, hawking, starobinsky1, bardeen1}, the action describing the evolution of a single real scalar field $\vf$ under the arbitrary potential $U = U(\vf)$ and minimally coupled to the metric $g_{\mu\nu}$ is
\begin{equation}
S = \int\,d^4x\,\sqrt{-g}\,\left[-\frac{1}{2}g^{\mu\nu}\,\frac{\partial \vf}{\partial x^\mu}\,\frac{\partial \vf}{\partial x^\nu} - U(\vf) \right],
\end{equation}
where $g \equiv {\rm det}(g_{\mu\nu}$). Here, we consider the mostly negative Friedmann-Robertson-Walker (FRW) metric. From the action $S$ follows the energy-momentum tensor
\begin{equation}
\mathcal{T}_{\mu\nu} = \frac{\partial \vf}{\partial x^\mu}\,\frac{\partial \vf}{\partial x^\nu} - g_{\mu\nu}\,\left[\frac{g^{\sigma\rho}}{2}\,\frac{\partial \vf}{\partial x^\sigma}\,\frac{\partial \vf}{\partial x^\rho}-U\right],
\end{equation}
with the energy density and pressure of the fluid associated to $\mathcal{T}_{\mu\nu}$ given respectively by
\begin{equation} \label{energy_pressure}
\rho = \frac{1}{2}g^{\mu\nu}\,\frac{\partial \vf}{\partial x^\mu}\,\frac{\partial \vf}{\partial x^\nu} + U,\quad p = \frac{1}{2}g^{\mu\nu}\,\frac{\partial \vf}{\partial x^\mu}\,\frac{\partial \vf}{\partial x^\nu} - U.
\end{equation}

Varying the inflaton field as $\vf(x) = \phi(x) + \delta\phi(x)$ and using the definitions in Eq.~\eqref{energy_pressure}, Eq.~\eqref{energy_conservation} gives
\begin{equation} \label{bkgd_motion_standard}
\ddf + 3H\,\df + U_\phi = 0,
\end{equation}
In scalar field theories, perturbations in the mode expansion $\delta \phi_k$ of the field operator,
\begin{equation} \label{fourier_transform}
\delta \phi(x) = \int\frac{d^3k}{(2\pi)^{3/2}}\,\left[\ca\,\delta \phi_k\,e^{i{\bf k}\cdot{\bf r}} + {\rm h.c.}\right],
\end{equation}
satisfy the differential equation
\begin{equation} \label{motion_standard}
\delta\ddot{\phi}_k + 3H\,\delta\dot{\phi}_k + \left(\frac{k^2}{a^2} + U_{\phi\phi}\right)\,\delta\phi_k = 0.
\end{equation}
During inflation, the Hubble rate in Eq.~\eqref{hubble_rate} is approximately constant due to the flatness of the potential $U$ and the fact that the kinetic term is negligible with respect to $U$. These conditions are equivalent to requiring that the two slow-roll parameters
\begin{equation} \label{slow_roll_condition}
\epsilon \equiv -\frac{\dot{H}}{H^2}, \quad \hbox{and} \quad \eta \equiv \frac{U_{\phi\phi}}{3H^2}.
\end{equation}
satisfy $\epsilon \ll 1$ and $|\eta| \ll 1$. The latter condition is equivalent to imposing $\ddot{\phi} \ll U_\phi$. In the standard cosmological theory, the slow-roll parameters in Eq.~\eqref{slow_roll_condition} read~\cite{liddle1993}
\begin{equation} \label{slow_roll_condition_standard}
\epsilon = \frac{1}{16 \pi\,G}\,\left(\frac{U_\phi}{U}\right)^2, \quad \hbox{and} \quad \eta = \frac{1}{8 \pi\,G}\,\frac{U_{\phi\phi}}{U}.
\end{equation}
Introducing $z = -1/aH$, setting $\delta \phi_k = z\,\chi_k$, and neglecting $U_{\phi\phi} = 3H^2\eta \ll H^2$, Eq.~\eqref{motion_standard} gives ($\chi_k' = d\chi_k/dz$)
\begin{equation} \label{motion_standard1}
\chi''_k + \left(k^2 - \frac{2+3(\epsilon-\eta)}{z^2}\right)\,\chi_k = 0,
\end{equation}
whose solution (for $\epsilon = \eta = 0$) corresponding to an incoming wave and reducing to a plane-wave for $kz\gg 1$~is
\begin{equation}
\chi_k(z) = \frac{e^{-ik\,z}}{\sqrt{2\,k}}\,\left(1 - \frac{i}{k\,z}\right).
\end{equation}
The inflaton field can thus be expanded in terms of lowering and raising operators $\ca$ and $\ca^{\dag}$ as
\begin{equation}
\hat{\chi}(z,{\bf r}) = \int\frac{d^3k}{(2\pi)^{3/2}}\,\left[\ca\,\chi_k(z)\,e^{i{\bf k}\cdot{\bf r}} + {\rm h.c.}\right].
\end{equation}
During inflation ($\ddf \ll H\,\df$), curvature perturbations generated on superhorizon scales are given by
\begin{equation}
\Rk = \frac{H\,\delta \rho_k}{\dot{\rho}} = \frac{U_\phi\,\delta\phi_k}{3(p+\rho)} = H\frac{\delta\phi_k}{\dot{\phi}}.
\end{equation}
The power spectrum of cosmological fluctuations is
\begin{equation} \label{spectrum_standard}
\Pk \equiv \langle\left|\mathcal{R}_k\right|^2\rangle = \frac{4\pi H^2}{|\dot{H}|}\,\Delta_\phi^2(k),
\end{equation}
where the fluctuations per logarithmic $k$ range is
\begin{equation}
\Delta_\phi^2(k) = \frac{k^3}{2\pi^2}\,\frac{|\chi_k(z)|^2}{a^2}.
\end{equation}
In the limit where $k\,|z| \ll 1$, which corresponds to the region where the wavelength is larger than the Hubble radius, we finally obtain the scale-invariant spectrum
\begin{equation} \label{spectrum_standard1}
\Delta_{\mathcal{R}}(k) = \frac{G\,H^4}{\pi\,|\dot{H}|} = \left(\frac{H^2}{2\pi\dot{\phi}}\right)^2.
\end{equation}
At the scale $k_* \!=\! 0.05{\rm Mpc}^{-1}$, the power spectrum is constrained at the 68\% confidence level (CL) as~\cite{ade2013} $\Pk = (2.215^{+0.032}_{-0.079})\times 10^{-9}$. The tensor-to-scalar ratio is constrained at 95\% CL as~\cite{ade2014}
\begin{equation} \label{tensortoscalar}
r = \frac{\PT}{\Pk} = \frac{16\,G\,H^2}{\pi\Pk} < 0.12.
\end{equation}
Writing $H^2 = \pi\,\Pk\,r/16G$ and $H^2 = (8\pi\,G/3)\,(U + \df^2/2)$, we derive
\begin{eqnarray}
\df &=& \frac{\sqrt{\Pk}\,r}{32G} = 2.2\times10^{31}{\rm~GeV^2}\,r_{0.1},\\ \nonumber
U \!&=&\! \frac{\Pk r (48\!-\!r)}{2048G^2} \!\approx\! \frac{3\Pk r}{128G^2} \!=\! 1.2\times 10^{65}r_{0.1}{\rm GeV^4},\\
\end{eqnarray}
where $r_{0.1} = r/0.1$. The scalar spectrum tilt is defined~as
\begin{equation} \label{spectral_tilt}
n_s - 1 \equiv \frac{d \ln \Pk}{d \ln k}\bigg|_{k=k_*},
\end{equation}
or, using the result in Eq.~\eqref{spectrum_standard1},
\begin{equation}
n_s - 1 = \frac{d}{d \ln k}\left[4\ln H - 2\ln \dot{\phi}\right]_{k=k_*} = -6\epsilon_* + 2\eta_*,
\end{equation}
where $\epsilon_*$ and $\eta_*$ are the slow-roll parameters at $k = k_*$.

\section{Viscous inflation model} \label{sec4}

A viscous term $\Pi = \Pi(t)$, possibly depending on time, may arise through either self-interaction~\cite{landau_book, huang_book} or the decay~\cite{zeldovich1970, murphy1973, hu1982, barrow1986, barrow1987, barrow1988, zimdahl} of the inflation field, acting as a negative pressure $p \to p-\Pi$~\cite{delcampo2007, delcampo2010, basterogil2011, basterogil2012, basterogil2014, visinelli2015}. In the FRW background, the equation of motion for the inflaton field reads
\begin{equation} \label{master_eq}
\dvf\,\left(\ddvf+ \frac{k^2}{a^2}\,\vf + U_{\vf} \right) + 3H\,\left(\dvf^2 - \Pi\right) = 0.
\end{equation}
Writing $\vf(x) = \phi(x) + \delta\phi(x)$ and introducing $\df = u$, the expression for the background component is obtained as
\begin{equation} \label{bkgd_motion_viscous}
\du + 3H\,u + U_\phi = 3H\frac{\Pi}{u},
\end{equation}
which reduces to the usual result in Eq.~\eqref{bkgd_motion_standard} for~$\Pi=~\!\!0$. Here instead, we show that Eq.~\eqref{bkgd_motion_viscous} with a vanishing potential $U=0$ also leads to an inflationary stage. We assume that viscosities are switched off whenever the momentum of the inflaton field is set to zero, or $\Pi = \zeta\,H = \zeta\,\lambda\,u$, where $\zeta$ is the viscosity coefficient, depending on $\phi$, and $H = \lambda\,u$ with $\lambda = \sqrt{4\pi G/3}$. We thus solve Eq.~\eqref{bkgd_motion_viscous} with $U = 0$,
\begin{equation} \label{bkgd_motion_viscous1}
\du + 3H\,u = \frac{3H^2\,\zeta}{u},
\end{equation}
or, using $\dot{u} = -Hzu'$ with $z=-1/aH$ and $u'=du/dz$,
\begin{equation} \label{eq_motion_viscous}
z\,u' - 3(u - \lambda\,\zeta) = 0.
\end{equation}
Assuming $\zeta = \zeta_0 + \zeta_1(z)$, where $\zeta_0$ is a constant and $\zeta_1(z)$ is a slowly-varying function, the solution to Eq.~\eqref{eq_motion_viscous} reads
\begin{equation} \label{viscous_solution}
u \!=\! \lambda\zeta_0 \!-\! \lambda\bz(z), \quad\hbox{with}\quad \bz(z) \!\equiv\! 3z^3\int^z\frac{\zeta_1(z')}{(z')^4}dz'.
\end{equation}
Using $\dot{H} = 4\pi\,G\left(\Pi - u^2\right) \approx 3\lambda^4\,\zeta_0\,\bz$ and introducing
\begin{equation}
\epsilon \equiv \frac{\dot{H}}{H^2} = \frac{3\bz}{\zeta_0},
\end{equation}
we obtain $\epsilon \ll 1$ for $\bz \ll \zeta_0/3$. A second constrain is obtained by using the slow-roll condition $\ddot{\phi} \ll 3H\dot{\phi}$ as
\begin{equation}
\du \ll 3Hu, \quad\hbox{or}\quad \beta \equiv \frac{z\,\bz'}{\zeta_0} = 3\frac{\zeta_1(z) + \bz(z)}{\zeta_0} \ll 1,
\end{equation}
where $\beta$ is a new slow-roll parameter and we used Eq.~\eqref{viscous_solution} to compute $\bz'$. Inflation ends when either one of these two conditions is no longer met, or $\epsilon(z_m),\beta(z_m) \sim1$, where $z_m =\exp(N_\phi)$ with a sufficient number of e-folds $N_\phi$ for successful inflation.

Using Eqs.~\eqref{spectrum_standard1} and~\eqref{tensortoscalar}, the scalar power spectrum and the tensor-to-scalar ratio in the viscous inflation are
\begin{equation} \label{spectrum_viscous}
\Pk = \frac{\lambda^6\,\zeta_0^3}{4 \pi^2\,\bz},\quad r = \frac{12\,\lambda^6\,\zeta_0^2}{\pi^2\Pk},
\end{equation}
from which we obtain
\begin{eqnarray} \label{viscous_term_u1}
\zeta_0 &=& \frac{\pi}{2\lambda^3}\,\sqrt{\frac{\Pk\,r}{3}} = 3.2\times 10^{50}\,r_{0.1}^{1/2}\,{\rm~GeV^3},\\
\bz &=& \frac{\pi}{96\lambda^3}\,\sqrt{\frac{\Pk\,r^3}{3}} = 6.6\times10^{47}\,r_{0.1}^{3/2}\,{\rm~GeV^3}.
\label{viscous_term_u2}
\end{eqnarray}
Notice that the tensor-to-scalar ratio and the first slow-roll parameter are related by $r = 16\epsilon$, as in the standard inflation model. The Hubble rate during inflation is
\begin{equation}
H_I = \lambda^2\,\zeta_0 = 3.7\times10^{13}{\rm~GeV}\,r_{0.1}^{1/2}.
\end{equation}

Using Eq.~\eqref{perturbation_energy_malik1}, we find that the Fourier transform of the scalar field perturbation $\delta\phi_k$ for the mode $k$, defined in Eq.~\eqref{fourier_transform}, satisfies the equation
\begin{equation} \label{motion_viscous}
\delta\ddot{\phi}_k \!+\! 3H\,\left(1 \!+\! \frac{\Pi}{u^2} \!-\! \frac{\Pi_u}{u}\right)\delta\dot{\phi}_k \!+\! \left(\frac{k^2}{a^2} \!-\! \frac{3H\,\Pi_\phi}{u}\right)\delta\phi_k \!=\! 0.
\end{equation}
Since $\Pi = \zeta\lambda\,u = \Pi_u\,u$ and
\begin{equation}
\Pi_\phi \equiv \frac{d\Pi}{d\phi} = -\frac{H\,z\,\Pi'}{u} = \frac{H\,\lambda^2\zeta_0^2}{u}\beta,
\end{equation}
Eq.~\eqref{motion_viscous} reads
\begin{equation} \label{motion_viscous1}
\delta\ddot{\phi}_k +3H\delta\dot{\phi}_k + \left(\frac{k^2}{a^2} - 3H^2\,\beta\right)\delta\phi_k = 0,
\end{equation}
which is the same expression as Eq.~\eqref{motion_standard}, with $\Pi_\phi$ playing the role of the potential curvature $U_{\phi\phi}$. Switching to the variable $z$ and writing $\delta \phi_k = z\,\chi_k$, we have
\begin{equation} \label{motion_standard1}
\chi''_k + \left(k^2 - \frac{2+3(\epsilon-\beta)}{z^2}\right)\,\chi_k = 0,
\end{equation}
with solution $\chi_k = C_1\sqrt{z}H^{(1)}(\nu,z)$, where $H^{(1)}(\nu,z)$ is the Hankel function of the first kind and
\begin{equation}
\nu = \frac{3}{2}\left(1+\frac{4}{3}(\epsilon-\beta)\right)^{1/2} \approx \frac{3}{2} + \epsilon-\beta.
\end{equation}
We obtain that the power spectrum predicted by the viscous inflation model is almost scale-invariant since, using Eq.~\eqref{spectral_tilt}, the spectral tilt results in
\begin{equation} \label{running_index}
n_s - 1 = -\frac{d \ln \Pk}{d \ln z}\bigg|_{z=z_*} = -4\epsilon_* + 2\beta_*.
\end{equation}
As long as the conditions for an accelerated expansion $\epsilon,\beta \ll 1$ are met, Eq.~\eqref{running_index} predicts a mild running of the power spectrum, while the viscous term $\zeta$ is approximately constant, see Eq.~\eqref{viscous_term_u1} with small corrections as in Eq.~\eqref{viscous_term_u2}. The exact value of $\zeta$ depends on the tensor-to-scalar ratio $r$, and it is calculable in models where the Hubble rate during inflation is predicted.

\begin{acknowledgments}
We acknowledge support by Katherine Freese through a grant from the Swedish Research Council (Contract No. 638-2013-8993).
\end{acknowledgments}

\end{document}